\documentclass{ws-ijqi}

\begin{document}

\markboth{Karl Svozil}
{Feasibility of the interlock protocol against man-in-the-middle attacks}

%
\catchline{}{}{}{}{}
%

\title{Feasibility of the interlock protocol against man-in-the-middle attacks on quantum cryptography}

\author{Karl Svozil}
\address{Institut f\"ur Theoretische Physik, University of Technology Vienna, \\
Wiedner Hauptstra\ss e 8-10/136, A-1040 Vienna, Austria                         \\
svozil@tuwien.ac.at}

\maketitle

\begin{abstract}
If an eavesdropper succeeds to compromise the quantum as well as the classical channels
and mimics the receiver ``Bob''  for the sender ``Alice'' and vice versa,
one defence strategy is the successive,
temporally interlocked partial transmission of the entire encrypted message.
\end{abstract}


\begin{quote}
\begin{flushright}
{\tiny
Ah, love, let us be true                                 \\
To one another! for the world, which seems                \\
To lie before us like a land of dreams,                  \\
So various, so beautiful, so new,                        \\
Hath really neither joy, nor love, nor light,            \\
Nor certitude, nor peace, nor help for pain;             \\
And we are here as on a darkling plain                   \\
Swept with confused alarms of struggle and flight,       \\
Where ignorant armies clash by night.                    \\
from {\it Dover Beach} by Matthew Arnold (1822-1888) }
\end{flushright}
\end{quote}

Since the introduction of conjugate coding by Wiesner
in the 70's of the last century \cite{wiesner},  quantum cryptography
\cite{benn-82,benn-84,ekert91,benn-92,gisin-qc-rmp,pflmubpskwjz}
has developed into an increasingly active area of applied research and technology,
as well as a powerful tool to exploit the quantum.
From the first realization at IBM's Yorktown Heights laboratory \cite{benn-92}
to its implementation across Lake Geneva \cite{gisin-qc-rmp} and
various other spots around the world, including
the Boston metropolitan area \cite{ell-co-05}
and the Viennese sewage system \cite{pflmubpskwjz},
quantum cryptographic techniques are among the finest experiments ever performed.

It is often argued that, as long as an eavesdropper ``Eve''
can only measure the communicated quanta,
the principles of physics and thus Nature itself protect the secrecy.
Several proofs of the unconditional security of quantum cryptographic protocols have been
published (e.g., Refs.~\cite{shpr-2000,may-2001}) supporting this line
of reasoning.
Few of them (e.g., Refs.~\cite{lue-98,lue-99,Gil-Ham-2000})
discuss man-in-the-middle attacks as a feasible cryptanalytic method.
Alas, while we have no intention to challenge these claims under the assumptions made,
not all eavesdroppers may stick to these rules.
It is not entirely unjustified to suspect that if
Eve is capable of compromising the quantum channel,
she may also be capable of compromising the classical channel.
Indeed, the possible interception of classical communication
is often taken as the very reason to propose an additional quantum channel
and quantum cryptographic protocols in general.
As either the security of the classical channel
or classical authentication is required for secure quantum cryptographic
protocols, it is not totally unjustified to state that
quantum cryptography cannot be considered qualitatively safer than classical cryptography.
It is nevertheless interesting to investigate the possible defence strategies against
cryptanalytic attacks. In what follows, after a brief review
of man-in-the-middle attacks, we shall concentrate on the interlock protocol as
an additional defence measure capable of reducing authentication.
Temporal interlocks have not yet been discussed in the quantum cryptographic context.

If Eve is able to intercept the classical as well as the quantum channels
between the sender Alice and the receiver Bob,
a rather straightforward man-in-the-middle attack can be launched~\cite{benn-84,benn-92,kuhn-03,pps-04,peev-04}
(see also the middleperson attack discussed in Ref.~\cite{BengioGrDeGoQu91}),
which has been discussed already in the BB84 paper~\cite{benn-84} but
seems to have gone unnoticed in the public perception of quantum cryptography
and is also not mentioned in many security proofs.

\begin{figure}[htbp]
  \centering
\unitlength 0.80mm
\linethickness{0.4pt}
\begin{picture}(155.00,45.00)
\put(24.98,9.98){\line(1,0){20.06}}
\put(34.96,12.48){\circle{4.99}}
\put(35.46,12.48){\circle{4.99}}
\put(24.98,9.68){\line(1,0){20.06}}
\put(24.98,19.96){\line(1,0){20.06}}
\put(34.96,22.46){\circle{4.99}}
\put(34.96,22.46){\makebox(0,0)[cc]{c}}
\put(35.06,12.48){\makebox(0,0)[cc]{q}}
\put(45.00,5.00){\dashbox{1.33}(20.00,20.00)[cc]{}}
\put(135.02,9.98){\line(-1,0){20.06}}
\put(125.04,12.48){\circle{4.99}}
\put(124.54,12.48){\circle{4.99}}
\put(135.02,9.68){\line(-1,0){20.06}}
\put(135.02,19.96){\line(-1,0){20.06}}
\put(125.04,22.46){\circle{4.99}}
\put(125.04,22.46){\makebox(0,0)[cc]{c}}
\put(124.94,12.48){\makebox(0,0)[cc]{q}}
\put(95.00,5.00){\dashbox{1.33}(20.00,20.00)[cc]{}}
\put(40.00,0.00){\framebox(80.00,30.00)[cc]{}}
\put(42.33,35.00){\makebox(0,0)[lc]{box-in-the-middle}}
\put(45.00,27.33){\makebox(0,0)[lc]{fake ``Bob''}}
\put(95.33,27.33){\makebox(0,0)[lc]{fake ``Alice''}}
\put(80.00,45.00){\makebox(0,0)[cc]{Eve}}
\put(135.00,5.00){\framebox(20.00,20.00)[cc]{}}
\put(135.33,27.33){\makebox(0,0)[lc]{Bob}}
\put(5.00,5.00){\framebox(20.00,20.00)[cc]{}}
\put(5.00,27.33){\makebox(0,0)[lc]{Alice}}
\put(65.00,20.00){\line(1,0){30.00}}
\put(80.00,20.00){\line(0,1){20.00}}
\put(80.00,15.00){\makebox(0,0)[ct]{copy or}}
\put(80.00,10.00){\makebox(0,0)[ct]{misinform}}
\end{picture}
 \caption{Scheme of a man-in-the-middle attack.}
\label{2004-bim-f1}
\end{figure}
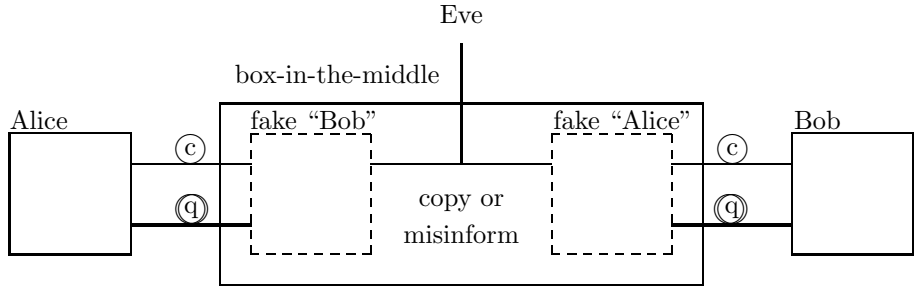

Let us thus shortly review man-in-the-middle attacks.
In the configuration discussed, we assume that a message is transferred from Alice to Bob,
and hence the scheme is forward directed.
In the case of key generation, like in the BB84 protocol, Eve's task is simpler,
as no indirect communication between Ann and Bob via Eve needs to take place;
Eve may communicate with each one of them separately.
Eve's part consists of three main phases.
The three phases are schematically represented in Fig.~\ref{2004-bim-f1}.
Very similar procedures hold for non-information-directed protocols used to generate keys.

\begin{itemize}
\item[(1)]
 In the first phase, Eve makes Alice believe that Alice is communicating with Bob
while she is actually communicating with Eve.
This, of course, can only be achieved by intercepting both classical and quantum channels.
Thereby, Alice's message which was intended for Bob can be deciphered and re-encoded classically.

\item[(2)]
 In the second phase, Eve processes the classical message to suit her goals.
For instance, the classical code can either be copied or modified.
Or, a totally new message, completely unrelated to Alice's original
message, may be prepared for the next phase.

\item[(3)]
 In the third phase, Eve invokes the same type of protocol as in the first phase
to re-encode the message prepared for Bob in the second phase,
and then sends it to Bob.
\end{itemize}

This attack is based on the problem of proper authenticating a legal sender and receiver.
Substitution of the classical channel by another quantum channel cannot solve the
authentication problem.

For a similar classical example, take
a message which is  encoded into two binary substrings $x$ and $y$
and transmitted over two separate classical channels.
Suppose a plain message which is to be transmitted secretly is coded into a binary sequence
$y=y_1y_2\ldots$, with $y_i\in \{0,1\}$ in any usual, unencrypted form.
Let the first channel convey an arbitrary random binary sequence $x=x_1x_2\ldots$,
with $x_i\in \{0,1\}$.
In the second channel, an encrypted sequence $z=z_1z_2\ldots$,
which is the sum modulus two
(the bitwise exclusive-or) $z_i=(x_i+y_i) \;{\rm  mod}\; 2$
of $x$ and $y$ is transmitted.
In such cases, an eavesdropper can only decipher the message and recover $y$ if both channels are intercepted.
The only difference between this scenario and the quantum one is
that, because of the no-cloning theorem, copying of a generic quantum bit is not allowed,
and hence $x$ and $z$, if at least one  of them
is communicated via a quantum channel,
cannot pass an eavesdropper unaltered.
But these obstacle can be circumvented by completely absorbing
and re-emitting a quantum code, as described above.

Usually, classical authentication is proposed as a defence
against man-in-the-middle attacks, resulting in a quantum cryptographic scheme which relies on
classical cryptology.
We shall consider here another strategy based on temporal synchronization.
This counter strategy cannot be directly applied to key generation,
as in this case no message is transferred between Alice and Bob.
But it may be helpful in an encoded information transfer between Alice and Bob.

Suppose Eve merely intends to copy the message and not to misinform.
Then, if the quantum protocol is not instantaneous,
only after completion of a sequence
she can start transmitting a copy of the message to Bob.
As a result, there is a time lag
during which Eve completes receiving Alice's message and retransmits it to Bob.

There are three shortcomings of a straightforward defence strategy based on time synchronization:
(1)
Time lags are intrinsic features of many quantum cryptographic protocols,
which do not result in instantaneous codes.
On the other hand, the bigger the time lags for decryption,
the easier it might be to detect eavesdropping through man-in-the-middle attacks.
(2)
Proper time synchronization between Alice and Bob would require a secure classical channel,
which was excluded by the assumptions.
However, for practical purposes, synchronization via satellite-based GPS systems
may be a feasible classical synchronization channel difficult to actively intercept.
The problem remains to securely negotiate the onset of the protocol,
as already in this phase Eve may negotiate with the respective partners.
(3)
In the case of misinformation (Eve misinforming Bob irrespective of the information she receives from Alice),
no copy of Alice's original message is required,
and there need not be any time lag at all.

\begin{figure}[htbp]
  \centering
\unitlength 0.8mm
\linethickness{0.4pt}
\begin{picture}(70.00,90.00)
\put(5.00,85.00){\makebox(0,0)[cc]{$t$}}
\put(66.33,5.00){\makebox(0,0)[cc]{$x$}}
\put(25.00,10.00){\line(0,1){80.00}}
\put(55.00,10.00){\line(0,1){80.00}}
\put(40.00,10.00){\line(0,1){80.00}}
\put(25.00,5.00){\makebox(0,0)[cc]{Alice}}
\put(40.00,5.00){\makebox(0,0)[cc]{Eve}}
\put(55.00,5.00){\makebox(0,0)[cc]{Bob}}
\put(25.00,15.00){\vector(1,1){30.00}}
\put(55.00,45.00){\vector(-1,1){30.00}}
\put(55.00,15.00){\vector(-1,1){30.00}}
\put(25.00,45.00){\vector(1,1){30.00}}
\put(10.00,10.00){\vector(1,0){60.00}}
\put(10.00,10.00){\vector(0,1){80.00}}
\put(5.00,15.00){\makebox(0,0)[cc]{(2)}}
\put(65.00,30.00){\makebox(0,0)[cl]{first part}}
\put(5.00,45.00){\makebox(0,0)[cc]{(3)}}
\put(65.00,60.00){\makebox(0,0)[cl]{second part}}
\put(5.00,75.00){\makebox(0,0)[cc]{(4)}}
\end{picture}
 \caption{Scheme of the interlock protocol. Numbers in brackets indicate the phases mentioned in the text.}
\label{2004-bim-f2}
\end{figure}
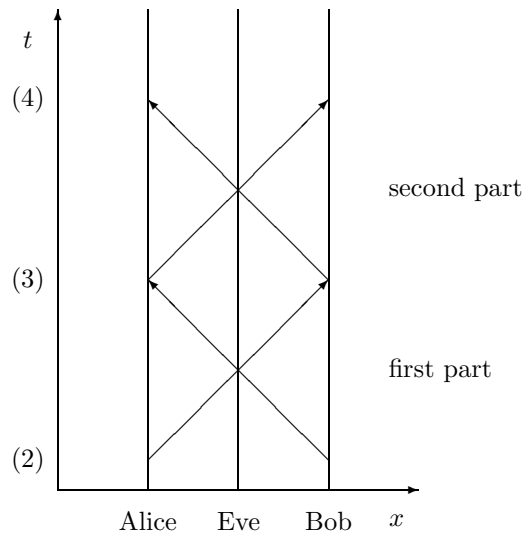

A refined strategy against classical man-in-the-middle attacks
is the interlock protocol proposed by Rivest and Shamir \cite{RivestSh84}
as well as timed-release cryptography by May \cite{May93,RivestSh96}.
The interlock protocol can be divided into four phases depicted in Fig.~\ref{2004-bim-f2}:
\begin{itemize}
\item[(1)]
Alice encrypts her message by a code which is not instantaneously decodable.
Conversely, Bob encrypts his message by a code which is not instantaneous.
(Part of this message may for instance consist of a classic authentication.)
\item[(2)]
Alice sends only a fraction of her encrypted message to Bob,
and Bob sends only a fraction of his encrypted message to Alice.
\item[(3)]
Upon receiving Bob's partial message,
Alice sends the rest of her encrypted message to Bob;
and upon receiving Alice's partial message, Bob sends the rest of his message to Alice.
(A generalization to the splitting of the encoded message into more than two packets is straightforward.)
\item[(4)]
Alice and
Bob put the two parts of the received encrypted messages together.
\end{itemize}

The advantage of the interlock over unbroken transmission
is that until phase three, Eve possesses only part of Alice's and Bob's encoded messages
and is unable to decode them for partial retransmission.
Only in the final phase, the entire messages can be recovered.
Thus, if time synchronization is maintained and observed, Eve has no other chance than to
fake the first messages exchanged by Alice and Bob.

The difference between classical and quantum interlock protocols resides
in the quantum no-cloning theorem and in complementarity:
in the quantum case, copying or measurement of the encoded messages exchanged by Alice and Bob is impossible
and would result in a randomization of these messages.
Absorption, decoding and the reemission of the re-encoded total message
is still possible, resulting in a time lag which could be monitored.
Alternatively, Eve could fake the packets of the first complete message exchanged
(thus misinforming Alice and Bob),
and could subsequently copy and retransmit the $(n-1)$'th packet in the duped communication
in the time slot reserved for a reception of the $n$'th packet.

Additional authentication of Alice and Bob would make it impossible for Eve to fake the
first messages exchanged.
Thus, if it would be possible, e.g., by using
satellite-based GPS systems, to synchronize Alice's and Bob's clocks,
then the interlock protocol could be applied as an additional resource
requiring classical authentication
to rule out a quantum man-in-the-middle attack
by demanding to keep the interlocked phases in temporal order.
As mentioned above, the ability to synchronize clocks represents
an additional authentication resource which is not present in the
quantum cryptographic protocols so far discussed or realized.

In summary,  sending interlocked partial sequences may be a practically feasible
defence against man-in-the-middle attacks on quantum encrypted message transfers.
It cannot be applied to quantum key generation.
Classical authentication methods remain indispensable to secure the legal identity of
the sender and the receiver.
With respect to man-in-the-middle attacks,
quantum cryptography shares the same vulnerability and cannot be considered principally safer
than classical cryptography.
We therefore agree with the position that quantum cryptography should be perceived more
as a secret key expansion \cite{benn-84,benn-82} or, in different words, a
secret key growing technique \cite{peev-04} rather than as a secret key generation scheme.
We believe that, instead of glorious claims, the vulnerabilities of quantum cryptography
need to be stated clearly for a proper comprehension
of the general public and also for the risk portfolio management of potential clients.


\end{document}